# Rabi oscillations in spasers during nonradiative plasmon excitation

E.S. Andrianov, A.A. Pukhov, A.V. Dorofeenko, A.P. Vinogradov
Institute for Theoretical and Applied Electromagnetics of the Russian Academy of Sciences
Izhorskaya st. 13, Moscow, 125412, Russia

A. A. Lisyansky
[2]Department of Physics, Queens College of the City University of New York, Flushing, NY
11367, USA

In the approach to the stationary regime, a spaser exhibits complicated and highly nonlinear dynamics with anharmonic oscillations.[1] We demonstrate that these oscillations are due to Rabi oscillations of the quantum dot in the field of the nanoparticle. We show that the oscillations may or may not arise dependent on the initial conditions.

## 1. INTRODUCTION

Recent discoveries in the area of nanoplasmonics, such as superlenses,[2] cloaking,[2,3] hyperlenses,[4-6] energy concentrators[7] and others, have raised high hopes for future development of ultrafast and super small optoelectronic devices. The extremely small localization length of surface plasmons (SPs), which are the basic modes of nanoplasmonic devices, allows one to overcome the fundamental limitation of the optical wavelength.

At optical frequencies, the physical principle of plasmonic devices operation is based on the plasmonic resonance of a metallic nanoparticle (NP). The main disadvantage of these materials is very large losses. To compensate for these losses, it was suggested using active (gain) inclusions.[6, 8-11] One of the most efficient and original type of an active inclusion is spaser (Surface Plasmon Amplification by Stimulated Emission of Radiation), which was first suggested in Ref. 12 and experimentally realized in Ref. 13. The spaser consists of a two-level quantum dot (QD) placed near a NP. The physical principle of the spaser's work is similar to the laser. The role of photons is played by SPs, which are localized at the NP.[1, 12, 14, 15] Confining SPs to the NP resembles a resonator. The spaser generates and amplifies the near field of the NP. The SP amplification occurs due to the nonradiative energy transfer from the QD to the NP. This process originates from the dipole-dipole (or any other near field[16]) interaction between the QD and the plasmonic NP. This physical mechanism has high efficiency because of the probability of the SP excitation is in $\left(kr\right)^{-3}$ larger than the probability of the radiative emission of the photon,[17] where $r$ is the distance between the centers of the NP and the QD, $k$ is an optical wavenumber in vacuum. The generation of a large number of SPs leads to the induced emission of the QD into the plasmonic mode and to the development of the generation of plasmons. Thus, the excitation of the plasmonic mode is provided by pumping through the excited QD. This process is inhibited by losses in the NP, which together with pumping results in stationary undamped oscillations of the spaser dipole moment. The key characteristic of spasers that determines their applicability in all-optical devices is the time of establishing of the stationary regime of a nanoplasmonic element.

It may seem that spasers cannot be used as amplifiers due to the independence of the final amplitude of their oscillations on initial values of the near fields.[18] However, recently Stockman[1] has suggested the possibility to design a spaser-based amplifier. Such an amplifier works during the transient regime, just after the population inversion has occurred but before the steady-flow regime in the spaser has been established. The transient regime takes about 250 fs[1] during which a spaser could amplify short optical pulses, making it an effective nanosize amplifier.

In the present paper, we study the dynamics of nonradiative plasmon excitations and investigate the emergence of SPs in a spaser and the conditions of establishing a steady state regime. On the way to the stationary regime, a spaser exhibits complicated and highly nonlinear dynamics with anharmonic oscillations.[1] We show that these oscillations are due to the Rabi oscillations of the QD in the field of the NP. The total time of the Rabi oscillations is smaller than the total transition regime time which is determined by the relaxation of the QD polarization. We also show the Rabi oscillations arise only for large initial values of the SP amplitude.

## 2. THE EQUATION OF THE SPASER DYNAMICS

Following Refs. [12, 19] let us consider the interaction between a metallic NP and a QD. For the simplest design of the spaser, it consists of a two-level QD of size $r_{TLS}$ located at a distance $r$ from the metal NP of size $r$ in a solid dielectric or semi-conductive medium. To describe the transition processes in such a spaser, one can use the model Hamiltonian of a metal NP interacting with a two-level QD in the form

$$\hat{H} = \hat{H}_{SP} + \hat{H}_{TLS} + \hat{V} + \hat{\Gamma}, \tag{1}$$

where

$$\hat{H}_{SP} = \hbar \omega_{SP} \hat{\tilde{a}}^\dagger \hat{\tilde{a}}, \tag{2a}$$

$$\hat{H}_{TLS} = \hbar \omega_{TLS} \hat{\tilde{\sigma}}^\dagger \hat{\tilde{\sigma}}. \tag{2b}$$

The Hamiltonians (2a) and (2b) describe the NP and the two-level QD, respectively,[12, 20, 21] their interaction is described by the operator $V = \hbar \Omega_R \sim (\mu_{NP} \mu_{TLS}) / r^3$, and the operator $\hat{\Gamma}$ describes all the effects of relaxation and pumping.[21] Below we take into account dipole SPs only. In Eqs. (2), $\hat{\tilde{a}}(t)$ is the Bose annihilation operator of the dipole SP, so that the electric field of the NP has the form

$$\hat{\mathbf{E}}(\mathbf{r}, t) = -A \nabla \varphi(\mathbf{r}) \left( \hat{\tilde{a}}^+ + \hat{\tilde{a}} \right),$$

where $A = \left( 4\pi\hbar s / \varepsilon_d s' \right)^{1/2}$, $s' = d \, \mathrm{Re}\left[ s(\omega) \right] / d\omega \big|_{\omega=\omega_n}$, $s = \left[ 1 - \varepsilon_{NP} / \varepsilon_M \right]^{-1}$ is the Bergman spectral parameter and $\varphi$ is the potential of the SP field, $\hat{\tilde{\sigma}} = |g\rangle\langle e|$ is the operator of the transition between the excited $|e\rangle$ and ground $|g\rangle$ states of the QD, $\hat{\boldsymbol{\mu}}_{TLS} = \boldsymbol{\mu}_{TLS} \left( \hat{\tilde{\sigma}}(t) + \hat{\tilde{\sigma}}^\dagger(t) \right)$ is the dipole moment of the QD, and $\boldsymbol{\mu}_{TLS}$ is its off-diagonal matrix element.

The SP wavelength, $\lambda_{SP}$, is much smaller than the optical wavelength in vacuum, $\lambda$, therefore the spatial derivatives in Maxwell equations are much larger than the temporal ones. Hence, the quasistatic approximation,[22] in which the temporal derivatives are neglected, can be used for the description of the SP field. For example, at the SP resonant frequency, a small spherical NP of radius $r_{NP} \ll \lambda$ turns out to be a half-length antenna (resonator), because the NP diameter is twice as small as the SP wavelength.[22] Thus, one can find the frequency of the plasmonic resonance from the condition of existence of the nontrivial solution of the Laplace equation

$$\nabla \varepsilon(\mathbf{r}) \nabla \varphi(\mathbf{r}) = 0, \tag{3a}$$

in which $\varepsilon(\mathbf{r})$ equals $\varepsilon_{NP}$ inside the NP and $\varepsilon_M$ outside the NP in the dielectric matrix. The value $\varepsilon_{NP}$ is considered as an eigenvalue of Eq. (3a). To separate the electromagnetic properties and the geometrical factor of the NP, it is convenient to represent the permittivity in the form[12] $\varepsilon(\mathbf{r}) = (\varepsilon_{NP} - \varepsilon_M)(\Theta(\mathbf{r}) - s)$, where the step function $\Theta(\mathbf{r})$ defines the geometry of the problem: it is equals to zero in the dielectric matrix and equals to unity inside the NP, whereas $s = [1 - \varepsilon_{NP} / \varepsilon_M]^{-1}$ only depends on constitutive properties of the system. By using such notification the Laplace equation may be written in the form[1, 14]

$$\nabla\Theta(\mathbf{r})\nabla\varphi(\mathbf{r}) = s\nabla^2\varphi(\mathbf{r}), \qquad (3b)$$

which allows one to find the eigenvalues $s_n$ and eigenfunctions $\varphi_n(\mathbf{r})$, $n = 1, 2, \ldots$ In the particular case of a spherical NP, the $n$-th plasmonic multipole has the resonance at $\varepsilon_{NP} = -\varepsilon_M(n+1)/n$ and $s_n = n/(2n+1)$[17, 20, 22].

By taking into account the permittivity dispersion, $\varepsilon_{NP}(\omega)$, the eigenfrequency $\Omega_n$ is determined from the condition $s(\Omega_n) = [1 - \varepsilon_{NP}(\Omega_n)/\varepsilon_M]^{-1} = s_n$. In the general case, due to the Joule losses in the metallic NP, the eigenfrequency has an imaginary part: $\Omega_n = \omega_n - i\gamma_n$. In addition, there are radiation losses of the NP due to the oscillation of a SP multipole. Let us estimate the values of both types of losses. In the case of the dipole SP, the part of losses associated with the emission of photons to free space can be estimated as

$$P_{rad} = \frac{2}{3c^3}\omega^4\mu_{NP}^2 = \frac{2}{27c^3}\omega^4|E|^2 r_{NP}^6(\varepsilon - 1)^2, \qquad (4)$$

where we take into account the relationship between the dipole moment with a homogeneous field inside the spherical NP and the external field: $\mu_{NP} = Er_{NP}^3(\varepsilon - 1)/3$. At the same time, the power absorbed due to Joule losses is

$$P_{Joule} = \frac{\omega}{8\pi}\int \varepsilon''|E|^2\,dV \approx \frac{\omega}{8\pi}\varepsilon''|E|^2\frac{4}{3}\pi r^3 = \frac{1}{6}r^3\omega\varepsilon''|E|^2. \qquad (5)$$

The ratio of these quantities is

$$\frac{P_{rad}}{P_{Joule}} = \frac{4}{9}\frac{(\varepsilon - 1)^2}{\varepsilon''}(k_0 r)^3. \qquad (6)$$

For the nanoparticle of size of 10 nm this ratio is ~0.06. Thus, for small nanoparticles Joule losses far exceed the loss due to radiation. This means that a spaser mainly generates the near field. However, for particles of size of 20 nm $P_{rad}/P_{Joule} \sim 0.5$ and the emission will be detected in far fields.

In this paper, we assume that the size of the NP is small and disregard the radiative losses. The imaginary part of the resonance frequency may be estimated as $\gamma_a = \omega \varepsilon'' \left[ \partial \left( \omega \varepsilon' \right) / \partial \omega \right]^{-1}$. For example, for a silver NP in the optical range $\gamma_a / \omega \sim 0.05$. We only consider the excitation of the main (dipole) mode with the frequency $\omega_{SP}$. The dipole moment of a spherical NP induced by the quasi-static field $\mathbf{E}_{TLS}$ of a QD with a dipole moment $\boldsymbol{\mu}_{TLS}$ equals to[23]

$$\boldsymbol{\mu}_{SP} = r_{NP}^3 \left( \frac{\varepsilon_{NP} - \varepsilon_M}{\varepsilon_{NP} + 2\varepsilon_M} \right) \mathbf{E}_{TLS} = \left( \frac{r_{NP}}{r} \right)^3 \left( \frac{\varepsilon_{NP} - \varepsilon_M}{\varepsilon_{NP} + 2\varepsilon_M} \right) \left[ \frac{3\left( \boldsymbol{\mu}_{TLS} \cdot \mathbf{r} \right) \mathbf{r}}{r^2} - \boldsymbol{\mu}_{TLS} \right], \tag{7}$$

Using the value of the permittivity for the silver NP surrounded by silicon oxide,[24] we estimate the dipole moment of the NP, $\mu_{SP}(\omega)$. The dipole moment of a typical QD of size $r_{TLS} \sim 10nm$ equals $\mu_{TLS} \approx 20D$.[25] Assuming $r_{NP} \sim r \sim 10nm$ we obtain $\mu_{NP} \sim 200D$ close to the plasmonic resonance. In turn, this allows one to estimate the energy of the static dipole-dipole interaction between the NP and the QD, $V = \hbar \Omega_R \sim \mu_{NP} \mu_{TLS} / r^3$. This gives $\Omega_R \approx 5 \cdot 10^{12} s^{-1} \ll \omega_{SP}$.

If the QD transition frequency is close to the frequency of SPs, $\omega_{SP} \approx \omega_{TLS}$, one can assume that the time dependence of operators $\hat{\tilde{a}}(t)$ and $\hat{\tilde{\sigma}}(t)$ has the form $\hat{\tilde{a}}(t) \equiv \hat{a}(t)e^{-i\omega t}$ and $\hat{\tilde{\sigma}}(t) \equiv \hat{\sigma}(t)e^{-i\omega t}$, where $\hat{a}(t)$ and $\hat{\sigma}(t)$ are the slowly varying amplitudes. Neglecting the fast oscillating terms $\sim e^{\pm 2i\omega t}$ (the rotating-wave approximation), the interaction operator can be written in the Jaynes-Cummings form[26]

$$\hat{V} = \hbar \Omega_R (\hat{a}^\dagger \hat{\sigma} + \hat{\sigma}^\dagger \hat{a}), \tag{8}$$

where $\Omega_R$ is the Rabi frequency. The commutation relations for operators $\hat{a}(t)$ and $\hat{\sigma}(t)$ are standard: $\left[ \hat{a}, \hat{a}^\dagger \right] = \hat{1}$ and $\left[ \hat{\sigma}^\dagger, \hat{\sigma} \right] = \hat{D}$, where $\hat{D}(t) = \hat{n}_e(t) - \hat{n}_g(t)$ is the population inversion operator, $\hat{n}_e = |e\rangle\langle e|$ and $\hat{n}_g = |g\rangle\langle g|$ are operators of the populations of the upper and lower levels of the QD, $\hat{n}_e + \hat{n}_g = 1$. Using Hamiltonian (1) we obtain the Heisenberg equations of motion for operators $\hat{a}(t)$, $\hat{\sigma}(t)$, and $\hat{D}(t)$:[14, 19]

$$\dot{\hat{D}} = 2i\Omega_R (\hat{a}^\dagger \hat{\sigma} - \hat{\sigma}^\dagger \hat{a}) - \frac{\hat{D} - \hat{D}_0}{\tau_D}, \tag{9}$$

$$\dot{\hat{\sigma}} = \left( i\delta - \frac{1}{\tau_\sigma} \right) \hat{\sigma} + i\Omega_R \hat{a}\hat{D}, \tag{10}$$

$$\dot{\hat{a}} = \left( i\Delta - \frac{1}{\tau_a} \right) \hat{a} - i\Omega_R \hat{\sigma}, \tag{11}$$

where $\delta = \omega - \omega_{TLS}$ and $\Delta = \omega - \omega_{SP}$ are frequency detunings. Since the time exponents in $\hat{n}_e = |e\rangle\langle e|$ and $\hat{n}_g = |g\rangle\langle g|$ are cancelled, the operator of the population inversion $\hat{D}(t)$ is slow. The time-constants $\tau_D$, $\tau_\sigma$ and $\tau_a$ are introduced in order to take into account the relaxation processes and the pumping term. The operator $\hat{D}_0$ describes pumping. The population inversion operator $\hat{D}_0$ plays the role of operator $\hat{D}$ in the regime of absence of generation. In other words, without generation $\{\hat{D} = \hat{D}_0, \hat{a} = \hat{\sigma} = \hat{0}\}$ is a stable fixed point of Eqs. (9)-(11). In the case when generation exists, this fix point becomes unstable (see below for details).[21, 26]

Below we neglect the quantum fluctuations and correlations and consider $\hat{D}(t)$, $\hat{\sigma}(t)$ and $\hat{a}(t)$ as complex quantities (*c*-numbers), substituting the Hermitian conjugation by the complex conjugation.[1, 11, 19, 23] $D(t)$ is a real valued quantity because the corresponding operator is Hermitian. The values of $\sigma(t)$ and $a(t)$ represent the complex amplitudes of the oscillations of the dipole moments of the QD and the NP, respectively.

## 3.   STATIONARY REGIME OF SPASER LASING

The system of equations (9)-(11) has a nontrivial stationary solution. Assuming that temporal derivatives equal zero, we reduce these equations to

$$2i\Omega_R (a^*\sigma - \sigma^* a) - \frac{D - D_0}{\tau_D} = 0, \tag{12}$$

$$\left( i\delta - \frac{1}{\tau_\sigma} \right)\sigma + i\Omega_R aD = 0, \tag{13}$$

$$\left( i\Delta - \frac{1}{\tau_a} \right)a - i\Omega_R \sigma = 0. \tag{14}$$

Eliminating $\sigma$ from Eqs. (13) and (14) we obtain

$$\left[ \frac{(i\delta - 1/\tau_\sigma)(i\Delta - 1/\tau_a)}{i\Omega_R} + i\Omega_R D \right] a = 0 . \tag{15}$$

Besides the trivial solution, $a = 0$, $\sigma = 0$ and $D = D_0$, there is a nontrivial solution corresponding to lasing. The complex expression in square brackets gives two real relations

$$\tau_\sigma \delta = -\tau_a \Delta , \tag{16}$$

$$D = D_{th} = \frac{1 + \Delta^2 \tau_a^2}{\Omega_R^2 \tau_a \tau_\sigma} . \tag{17}$$

The lasing frequency is given by Eq. (16):

$$\omega = \frac{\omega_{SP}\tau_a + \omega_{TLS}\tau_\sigma}{\tau_a + \tau_\sigma} . \tag{18}$$

The frequency defined by Eq. (18) is always located between frequencies of NP and QD.[12, 19] From Eqs. (8)-(10) and (13) we find the stationary amplitude of NP and QD dipole moments:

$$a = \frac{e^{i\psi}}{2} \sqrt{\frac{(D_0 - D_{th})\tau_a}{\tau_D}} , \tag{19}$$

$$\sigma = a\frac{\Delta + i/\tau_a}{\Omega_R} , \tag{20}$$

where $\psi$ is an arbitrary constant phase. Because $D \le D_0 \le 1$, the nontrivial stationary solution has physical meaning only if $D_{th} < 1$ or

$$\Omega_R^2 \tau_a \tau_\sigma > 1 + \Delta^2 \tau_a^2 . \tag{21}$$

Eq. (21) is a necessary condition for spasing: if this condition is satisfied and $D_0 > D_{th}$, the nontrivial stationary solutions (17), (19) and (20) exist. This means that non-zero dipole moments of the NP and the QD are excited, i.e. the spaser generates plasmons. For the parameters used in our consideration, $D_{th} \sim 0.1$. This value is small enough to be achieved experimentally.

## 4. TRANSIENT REGIME

The system of equations (9)-(11) is nonlinear, therefore the establishing of the steady-state regime can be investigated only numerically. Our computer simulation shows that the system evolution strongly depends on initial conditions and on the relations between relaxation times $\tau_a$, $\tau_\sigma$ and $\tau_D$ (see also Ref. 1). As one can see from Fig. 1, this evolution can be either monotonic or have an oscillatory character.

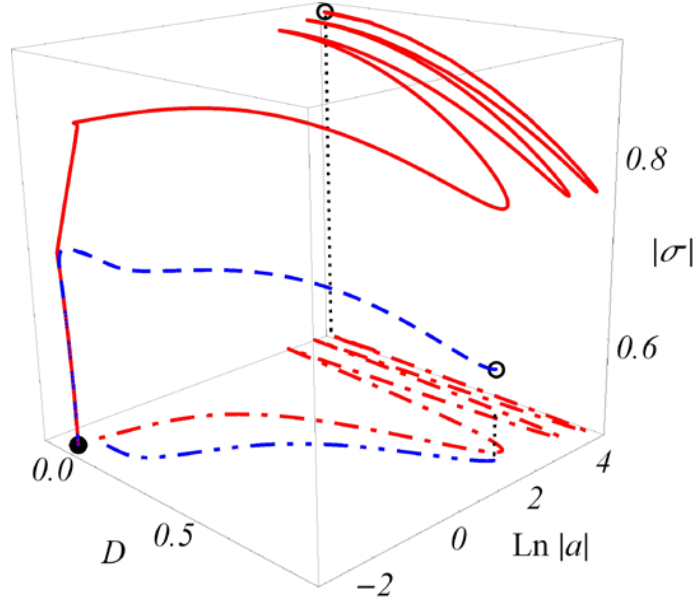

Fig. 1. The evolution of the spaser. The open circles mark the beginning of the transient regime, then the system moves along either the dashed ($a(0) = 40 + 25i$, $\sigma(0) = 0.9$ and $D(0) = 0.05$) or the dashed line ($a(0) = 5$, $\sigma(0) = 0.65$ and $D(0) = 0.9$) until it reaches the stationary state marked by the solid circle. Dash-dotted and dash-double dotted lines are projections of the trajectories on the plane $|\sigma| = 0.5$. The curves shown were calculated for the time-constants $\tau_a = 10^{-14} s$, $\tau_\sigma = 10^{-11} s$, $\tau_D = 10^{-13} s$, and $\Omega_R = 10^{13} s^{-1}$.

The dynamics of the transition regime is characterized by the energy exchange between the SP of the NP and the QD excitation. The phase difference $\Delta\varphi(t)$ between the QD polarization and the SP ($\sigma(t) = e^{i\Delta\varphi} a(t)$) is responsible for the energy flux.[27] In particular, the sign of $\sin\Delta\varphi$ determines the direction of the flux. Thus, oscillations of $\sin\Delta\varphi$ indicate the change in the direction of the energy flux.

The time dependence of $\Delta\varphi$ is sensitive to the initial values of $a$, $\sigma$ and $D$. Note, that the latter two quantities according to their physical meanings should be chosen within the intervals $0 \le \sigma(0) \le 1$ and $D \le D_0$. For small initial values of the SP amplitude, $a(0) \ll 1$, the system parameters do not oscillate. The transition time is determined by the relaxation time $\tau_\sigma$.

The value of $\sin\Delta\varphi(t)$ increases monotonically from the initial value, which is equal to zero, to the steady-state value that corresponds to the energy flux directed from the QD to the NP (the dashed line in Fig. 1 and the solid line in Fig. 2).

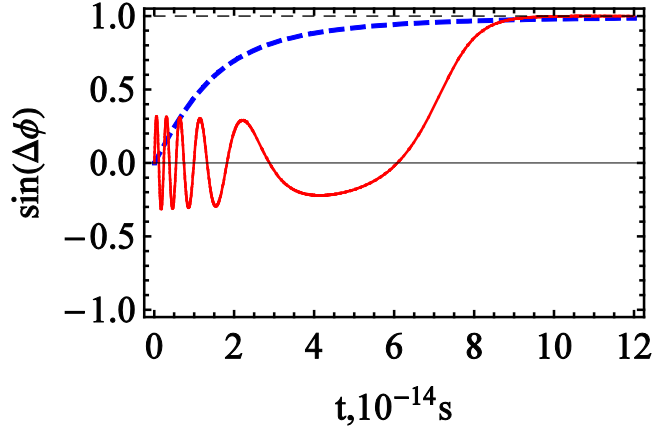

Fig. 2. (Color online) The dependence of the phase mismatch on time. The initial conditions corresponding to solid line and dashed lines are $a(0) = 50i$, $\sigma(0) = 0.25$, $D(0) = 0.6$ and $a(0) = 0.1$, $\sigma(0) = 0.5$, $D(0) = 0.7$, respectively. Values of the time constants and the Rabi frequency are the same as in Fig. 1.

For large initial values of the SP amplitude, $a(0) \gg 1$, the variables $a$, $\sigma$ and $D$ exhibit oscillations as shown in Figs. 1 and 3. In this case, the transient regime splits into two stages. At the first stage, which lasts $\sim \tau_a \ln|a(0)|$, there are the Rabi oscillations with the characteristic period $\sim \pi / \Omega_R |a(0)|$. The total number of oscillations $N$ depends on the initial plasmon amplitude, $N \sim \tau_a \Omega_R |a(0)| \ln|a(0)|$. During the second stage, the QD polarization gradually reaches the stationary value without oscillations (see Fig. 3).

At the first stage, $\sin\Delta\varphi(t)$ oscillates (the solid line in Fig. 2). This indicates that the energy flux changes its direction. Thus, the energy can flow not only from the QD to the NP but also from the NP to the QD.

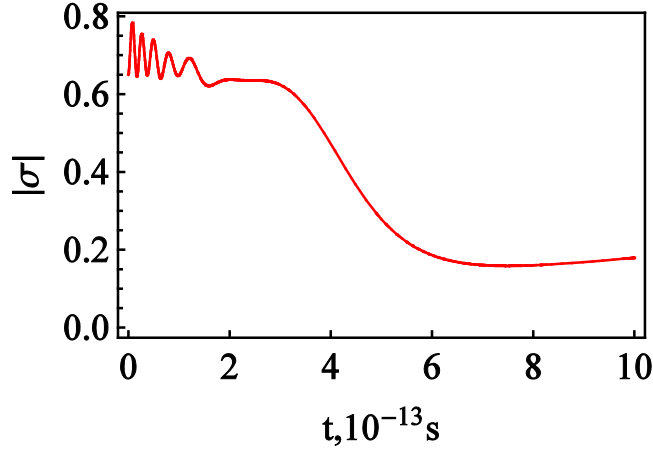

Fig. 3. The dependence of the QD polarization on time calculated for the initial conditions $a(0) = 10$, $\sigma(0) = 0.65$ and $D(0) = 0.2$. All other parameters have the same values as in previous figures.

## 5.  RABI OSCILLATIONS OF NONRADIATIVE SPASER EXCITATIONS

To estimate the frequency of the oscillations observed in the computer simulation,[12] for simplicity, we consider the case of the exact resonance $\delta = \Delta = 0$. Then the system of equations (9)-(11) has the form

$$\dot{a} = -i\Omega_R \sigma \,, \tag{22}$$

$$\dot{\sigma} = i\Omega_R aD \,, \tag{23}$$

$$\dot{D} = 2i\Omega_R (a^*\sigma - \sigma^* a) \,. \tag{24}$$

During the first stage, when all dynamic variables of the spaser are far from their stationary values, the time derivatives are much greater than respective terms proportional to the inverse relaxation times, therefore we can omit the terms responsible for the relaxation and pumping. The substitution of Eq. (22) and its conjugate into Eq. (24) gives

$$\dot{D} = -2\left(a^*\dot{a} + \dot{a}^*a\right) = -2\frac{d|a|^2}{dt} \,. \tag{25}$$

This leads to the following integral of motion

$$D + 2|a|^2 = C_1 \,, \tag{26}$$

where $C_1 = D(0) + 2|a(0)|^2$. Representing $a$ in the form $a = |a|e^{i\varphi}$ and differentiating Eq. (22)

with respect to time, we obtain

$$\left[ \frac{d^2|a|}{dt^2} - |a|\left(\frac{d\varphi}{dt}\right)^2 + i\left(2\frac{d\varphi}{dt}\frac{d|a|}{dt} + |a|\frac{d^2\varphi}{dt^2}\right)\right]e^{i\varphi} = i\Omega_R\frac{d\sigma}{dt}. \tag{27}$$

Using Eqs. (23), (25) and (27), we can split Eq. (27) into two equations

$$2\frac{d\varphi}{dt}\left|\frac{da}{dt}\right| + |a|\frac{d^2\varphi}{dt^2} = 0, \tag{28}$$

$$\frac{d^2|a|}{dt^2} - |a|\left(\frac{d\varphi}{dt}\right)^2 = \Omega_R^2|a|(C_1 - 2|a|^2). \tag{29}$$

From Eq. (28) it is clear that

$$|a|\left(2\frac{d\varphi}{dt}\left|\frac{da}{dt}\right| + |a|\frac{d^2\varphi}{dt^2}\right) = \frac{d}{dt}\left(|a|^2\frac{d\varphi}{dt}\right) = 0. \tag{30}$$

As the result, we obtain the second integral of motion $|a|^2\left(d\varphi/dt\right) = C_2$. Let us suppose that $C_2 = 0$ that can always be achieved by the appropriate choice of initial conditions. Then $d\varphi/dt = 0$ and from Eq. (29) we obtain

$$\frac{d^2|a|}{dt^2} = \Omega_R^2|a|(C_1 - 2|a|^2). \tag{31}$$

Eq. (31) represents Newton's equation for a unit mass particle with the coordinate $|a|$ moving in the potential $U\left(|a|\right) = \left(\Omega_R^2|a|^4 - C_1\Omega_R^2|a|^2\right)\!/2$. The stable equilibrium position of such a particle is $|a|_{eq} = \sqrt{C_1/2} = \sqrt{|a(0)|^2 + D(0)/2} \approx |a(0)|$. Thus, the particle exhibits oscillations about this equilibrium position with the frequency

$$\Omega = 2|a(0)|\Omega_R. \tag{32}$$

This expression coincides with the known expression for the frequency of Rabi oscillations, which occur for the interaction of a two-level QD with the classical harmonic field of the amplitude $a(0)$ or the quantized field with the number of quanta $\hat{a}^+(0)\hat{a}(0) = n = |a(0)|^2$.[26] In addition, Eq. (32) is in a full agreement with the simulation results. Therefore, we can conclude that the phase oscillations observed in the numerical simulation are Rabi oscillations of QD populations in the NP near field.

# 6.  DISCUSSION OF RESULTS

Our numerical simulation shows that in the stationary regime, similar to the case of the interaction of two classical dipoles,[27] the excitation of the NP by the QD near field is possible only if there is the phase difference, $\Delta\varphi$, between the dipole moments of the QD and the NP. When $\sin\Delta\varphi$ is positive, the energy provided by the pump in the QD is transmitted to the NP, and then is dissipated in the form of loss. During the transition regime the direction of the energy flow may change several times.

The spaser transition to the stationary auto-oscillations is determined by three characteristic times: the relaxation time of the SP amplitude, $\tau_a$, the relaxation time of the QD polarization, and the relaxation time of the population inversion, $\tau_D$. $\tau_a$ stems from Joule losses in metal NP. Due to extremely high losses in metals this time is the shortest. Its experimentally measured value is $10^{-14}$–$10^{-13}$ s[28] that coincides with estimates obtained from classical electrodynamics.[9] The typical values of $\tau_\sigma$ and $\tau_D$ known from experiments are $10^{-11}$ s and $10^{-13}$ s, respectively.[29-31] Thus, for metal NPs and semiconductor nanocrystal QDs, typical relations for times $\tau_a$, $\tau_D$, and $\tau_\sigma$ are $\tau_a < \tau_D \ll \tau_\sigma$. The total transition is determined by the largest time $\tau_\sigma$.

The character of the transient process is greatly affected by the initial value of the SP amplitude $a(0)$. For the "cold start" of the spaser, when the initial value of SPs is due to the spontaneous emission of a QD, $a(0)$ is small. Then the electric field of the SP is of the order of the field of the dipole moment of the QD and the Rabi oscillations do not arise and the energy flows monotonically from the QD to the NP.  A large value of $a(0)$ can be obtained by exciting SPs with a nanosecond pulse of an optical parametric oscillator.[12] In this case, the transition process is more complicated. It can be divided into two stages. During the first stage, when the QD is in the high field of the SP, the Rabi oscillations with the characteristic period $\tau_R = \Omega_R^{-1}$ arise (see Fig. 3) and the energy flux oscillates between the QD and the SP. During the time $\sim \tau_a \ln|a(0)|$ these oscillations die out due to dissipation. During the second stage, the spaser exhibits a behavior observed at small $a(0)$, the transient regime continues until the spaser starts stationary auto-oscillations. The total time of the transient regime is $\sim \tau_\sigma$. For $\tau_a < \tau_D \ll \tau_\sigma$ the picture depends weakly on the value of $\tau_D$.

During the period in which stationary spaser oscillations are established, the spaser can be used as an amplifier or a switch. The total duration of the transition regime is determined by the relaxation of the QD polarization, $\tau_\sigma$. Thus, one should use QDs with slow relaxation in designing a nano-amplifier, while using QDs with fast relaxation is preferable for ultrafast switching.

## ACKNOWLEDGEMENTS

The authors are indebted to B. Luk'yanchuk for useful discussions. This work was supported by RFBR Grants Nos. 10-02-91750, 10-02-92115 and 11-02-92475 and by a PSC-CUNY grant.